\documentclass[runningheads]{llncs}
\usepackage{graphicx}
\usepackage{booktabs}

\usepackage{amsthm}
\usepackage[hidelinks]{hyperref}
\begin{document}
\title{HyperSec: Visual Analytics for blockchain security monitoring}
\author{Benedikt Putz\inst{1}\orcidID{0000-0002-4265-1106} \and
Fabian Böhm\inst{1}\orcidID{0000-0002-0023-6051} \and
Günther Pernul\inst{1}}
\authorrunning{B. Putz et al.}
\institute{University of Regensburg, Regensburg, Germany\\
\email{\{firstname.lastname\}@ur.de}}
\maketitle              % typeset the header of the contribution
\begin{abstract}
Today, permissioned blockchains are being adopted by large organizations for business critical operations. Consequently, they are subject to attacks by malicious actors. Researchers have discovered and enumerated a number of attacks that could threaten availability, integrity and confidentiality of blockchain data. However, currently it remains difficult to detect these attacks. We argue that security experts need appropriate visualizations to assist them in detecting attacks on blockchain networks. To achieve this, we develop HyperSec, a visual analytics monitoring tool that provides relevant information at a glance to detect ongoing attacks on Hyperledger Fabric. For evaluation, we connect the HyperSec prototype to a Hyperledger Fabric test network. The results show that common attacks on Fabric can be detected by a security expert using HyperSec's visualizations.%, without accessing a command line.

\end{abstract}
%

%%%%%%%%%%%%%%%%%%%%%%%%%%%%%%%%%%%%%%% INTRODUCTION %%%%%%%%%%%%%%%%%%%%%%%%%%%%%%%%%%%%%%%
{\let\thefootnote\relax\footnote{{The first two authors have contributed equally to this manuscript.}}}
\section{Introduction}
    %Motivation
    New use cases of distributed ledger technology (DLT) are proposed on a daily basis by academia and practice, leading to an increasing number of projects and solutions. Beyond that, blockchain applications are increasingly being used in real large-scale supply chain environments, such as the TradeLens \cite{Jensen2019} and DLFreight \cite{DLTLabs2020} platforms. At first glance, DLT seems to increase an application's security or even solve existing applications' security issues. However, the task of securing the DLT itself is often neglected in practice due to its complexity and the number of serious challenges connected to it.
    
    % Challenges
    The complexity of blockchain technology makes it particularly challenging to identify malicious activities \cite{Boshmaf2019}. In any blockchain network, there are several independent peers operated by independent organizations, where each organization only has a limited view of the network. Each node also has various data sources from its components, making it difficult to obtain an overview of the network's state \cite{Putz2020}. Since blockchain is a networked database, it also requires monitoring both the host and the network, which results in a large volume and velocity of observable data.  
    
    Fully automated systems for attack detection on blockchains do not yet exist. Even if respective technologies for blockchain security monitoring were available, human experts remain indispensable as their domain knowledge is crucial to identify and analyze intricate attack patterns \cite{BenAsher2015}. Therefore, we need a way to make the heterogeneous data at hand available for domain experts. Visualizations offer a well-known path to achieve this goal. A visual representation can help a domain expert make sense of the displayed information and efficiently draw conclusions \cite{Keim2008}. These observations lead to our work's research question: 
    
    \newtheorem*{researchquestion}{RQ}
    \begin{researchquestion}
        What are appropriate visualizations to assist security experts in detecting DLT threats?
    \end{researchquestion}
    
    %Contribution
    In this work, we make a two-fold contribution to this research question. We first formulate the domain problem of monitoring permissioned DLTs for threats or ongoing attacks. Structuring the domain problem and deriving general design requirements serves as the foundation for our visualization approach. The second part of our contribution is the task-centered design and prototypical implementation of \textit{HyperSec}, a visual representation of security-relevant DLT information to support security experts' monitoring tasks.

    %Remainder
    The remainder of this work is structured as follows. Section \ref{sec:related} gives a brief overview of related academic work in the field of security visualizations in the blockchain domain. In Section \ref{sec:blockchain_security_monitoring}, we flesh out the domain problem faced by security experts monitoring permissioned blockchain environments for immediate threats. Section \ref{sec:hypersec} then introduces our visualization design and its prototypical implementation using open source technologies. Afterward, we evaluate our visualization design by simulating attacks in Section \ref{sec:use_case}. Finally, Section \ref{sec:conclusion} concludes our work with a summary and possible future research directions.

%%%%%%%%%%%%%%%%%%%%%%%%%%%%%%%%%%%%%%% RELATED WORK %%%%%%%%%%%%%%%%%%%%%%%%%%%%%%%%%%%%%%%
\section{Related Work}
\label{sec:related}
    Recently, Tovanich et al.~\cite{Tovanich2019} carried out a systematic review to structure existing work on the visualization of blockchain data. Their research and previously conducted studies~\cite{Sundara2017} identify several visualization approaches with a focus on criminal and malicious activity~\cite{DiBattista2015,McGinn2016}. Although these surveys highlight that visualization tools for blockchains are on the rise, two significant shortcomings concerning security monitoring become apparent. First of all, most of these tools focus on public blockchains (i.e., cryptocurrencies) and are not very well adaptable to private blockchains~\cite{Yli-Huumo2016}. Secondly, existing visualization approaches to analyze criminal or malicious activities within a blockchain network focus on detecting the events only after they have occured~\cite{Tovanich2019}. 
    
    To effectively prevent attacks upfront, blockchain networks have to be actively monitored by blockchain security experts. Several studies discuss external and internal threats that could impair a blockchain network's functionality \cite{Homoliak2020,Putz2019Insider}. Zheng et al. propose a framework for monitoring the Ethereum blockchain's performance \cite{Zheng2018}. They introduce some respective metrics while using both node logs and Remote Procedure Calls (RPC) to gather data. Threat indicators to detect malicious activities in a blockchain network have recently been introduced by Putz et al.~\cite{Putz2020}. Based on this limited body of work from academia, blockchain metrics and threat indicators need to be made available to security experts for effective monitoring. Existing monitoring solutions like the dashboard by Bogner~\cite{Bogner2017} focus only on transaction activity but do not consider other security-relevant data and metrics.
    
    An approach pointing in this direction is the Hyperledger Explorer\footnote{\url{https://www.hyperledger.org/use/explorer}}, the Hyperledger project's tool for monitoring Hyperledger blockchains. The Explorer connects to a local blockchain node and extracts data about blocks, transactions, peers, and more into a local PostgreSQL database. %The extracted data can be accessed using a REST API. 
    Additionally, a web application is available for inspecting blockchain data, including some basic visualizations of transaction data. However, these visualizations are not tailored to provide the necessary insights or indicators to detect threats. %Another Hyperledger Labs project integrates Fabric with ElasticSearch and Kibana, resulting in a Kibana dashboard able to display some transaction data \footnote{\url{https://github.com/hyperledger-labs/blockchain-analyzer/}}. Similar to Hyperledger Explorer, only the nodeJS SDK is leveraged to download and index transaction data in ElasticSearch. The Kibana dashboard itself is not targeted at a specific task (like security monitoring).
    
    Analyzing related work highlights an evident lack of dedicated and security-specific visualization approaches enabling security experts to monitor blockchain networks and detect common indicators of compromise or ongoing attacks on the network. Our work contributes a valuable solution approach to this issue.

%%%%%%%%%%%%%%%%%%%%%%%%%%%%%%%%%%%%%%% DOMAIN PROBLEM %%%%%%%%%%%%%%%%%%%%%%%%%%%%%%%%%%%%%%%
\section{Blockchain Security Monitoring}
\label{sec:blockchain_security_monitoring}
    This Section addresses the first part of our contribution. Within our main contribution, we follow the user-centered and problem-driven Nested Blocks and Guidelines model (NBGM) for visualization designs \cite{Meyer2015,Munzner2009}. This allows us to identify and address security experts' core problems and lay a foundation for a visualization design fitting their needs. 

    The first step of the NBGM is the definition of a domain problem. We characterize the problem at hand based on two primary sources of information. First, we consider reports from blockchain security professionals \cite{Kacherginsky2020}. Second, we analyze literature on blockchain attacks to identify concerns for operators of a blockchain node \cite{Dabholkar2019,Homoliak2020,Putz2019Insider}. We begin by outlining the overall blockchain security monitoring process in Section \ref{subsec:process}. The domain problem is then specified according to Miksch and Aigner's design triangle through more in-depth descriptions of specific users (Section \ref{subsec:users}), their tasks (Section \ref{subsec:tasks}), and data elements (Section \ref{subsec:data}) \cite{Miksch2014}. We address the second step of the NBGM (\textit{Data/Operation Abstraction}) in Section \ref{subsec:design_requirements} by deriving general design requirements for a visualization approach to support blockchain security monitoring.
    
    \subsection{Blockchain Security Monitoring Process}
    \label{subsec:process}
\begin{figure}[t]
        \centering
        \includegraphics[width=\linewidth]{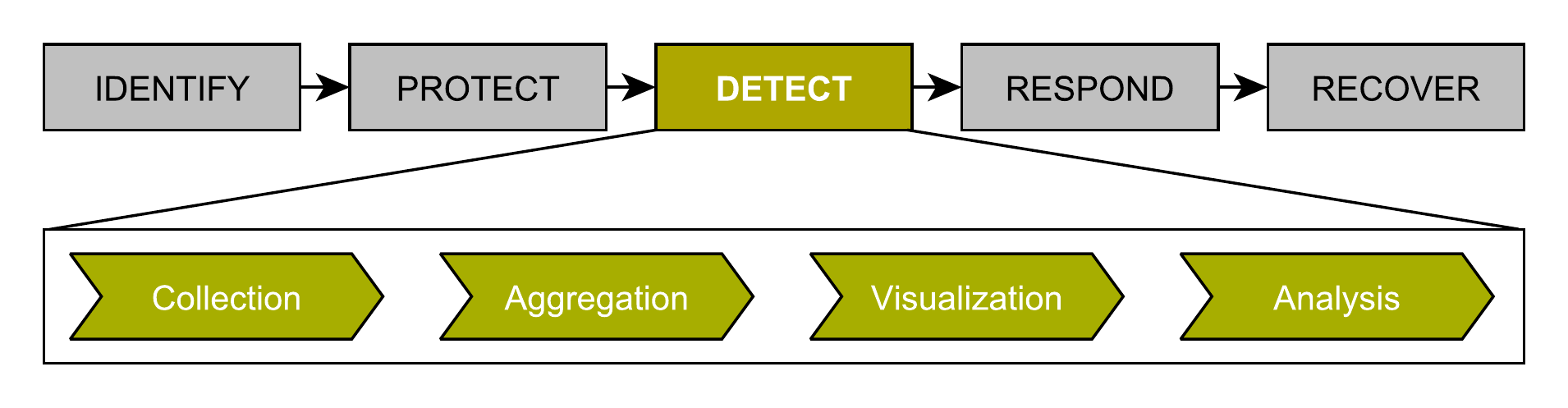}
        \caption{Blockchain Security Monitoring process based on the NIST Cybersecurity Framework \cite{Calder2018}.}
        \label{fig:process}
    \end{figure}

    Before we dive into users, tasks, and available data, we first need to understand the overall process underlying blockchain security monitoring. This subsection introduces our conceptual process based on the NIST Cybersecurity Framework for protecting critical infrastructures \cite{Calder2018}. As shown in Figure \ref{fig:process}, the framework has five main functions: \textit{Identify}, \textit{Protect}, \textit{Detect}, \textit{Respond} and \textit{Recover}. We apply these functions to a permissioned blockchain network. The \textit{Identify} function serves to identify relevant assets and risks. This problem has been already addressed in prior work \cite{Putz2020}. \textit{Protect} involves a variety of protection measures applied to the system: identity management and access control, data security, secure configuration, and backups/log files, among others. These protection measures are usually part of the blockchain framework itself, with additional measures being applied at deployment time (such as secure configuration and appropriate backup procedures) \cite{TheLinuxFoundation2020a}. The \textit{Detect} function currently lacks appropriate visualization and analysis tools. It's the focus of this work and further developed in the following subsection. During the \textit{Respond} phase, threats detected using our visualization approach are met with a response plan and appropriate mitigation actions. Finally, the \textit{Recover} function provides appropriate tools to restore functionality after an attack has occurred. \textit{Respond} and \textit{Recover} are not specifically part of this work as attacks need to be identified before effective \textit{Respond} and \textit{Recover} can take place. Corresponding tools might be integrated into future work to permit swift threat response.
    
    The \textit{Detect} function can be subdivided into four smaller process steps. Relevant data needs to be collected (\textit{Collection}) and aggregated to provide appropriate metrics if necessary (\textit{Aggregation}). Data and metrics can then be visualized (\textit{Visualization}) allowing domain experts to identify possible threats (\textit{Analysis}). Please note that all steps beside \textit{Analysis} can be performed automatically.
    
    \subsection{Users}
    \label{subsec:users}
    The intended users of visualization designs within the \textit{Detect} function in the blockchain monitoring process are domain experts. These experts are responsible for analyzing blockchain data to identify malicious events within this function \cite{Kacherginsky2020}. More specifically, we define the domain experts as security professionals knowledgeable in the cybersecurity domain. Therefore, we expect them to have the expertise to decide whether specific events or event series indicate an imminent threat to the blockchain. Within a permissioned blockchain, these security experts are responsible for monitoring the distributed network through the view of the local organization's blockchain node. Other organization's nodes could also be monitored, but data availability is likely limited due to access restrictions within the blockchain network. 
    
    \subsection{Tasks}
    \label{subsec:tasks}
    Visualizations or any other tool supporting the \textit{Detect} function of the blockchain security monitoring process should be based on the tasks that the respective users need to carry out. Following the user characterization above, we derive the crucial tasks of the domain expert's work.
    
    \begin{figure}[t]
        \centering
        \includegraphics[trim={1cm 1cm 0cm 0cm},width=\linewidth]{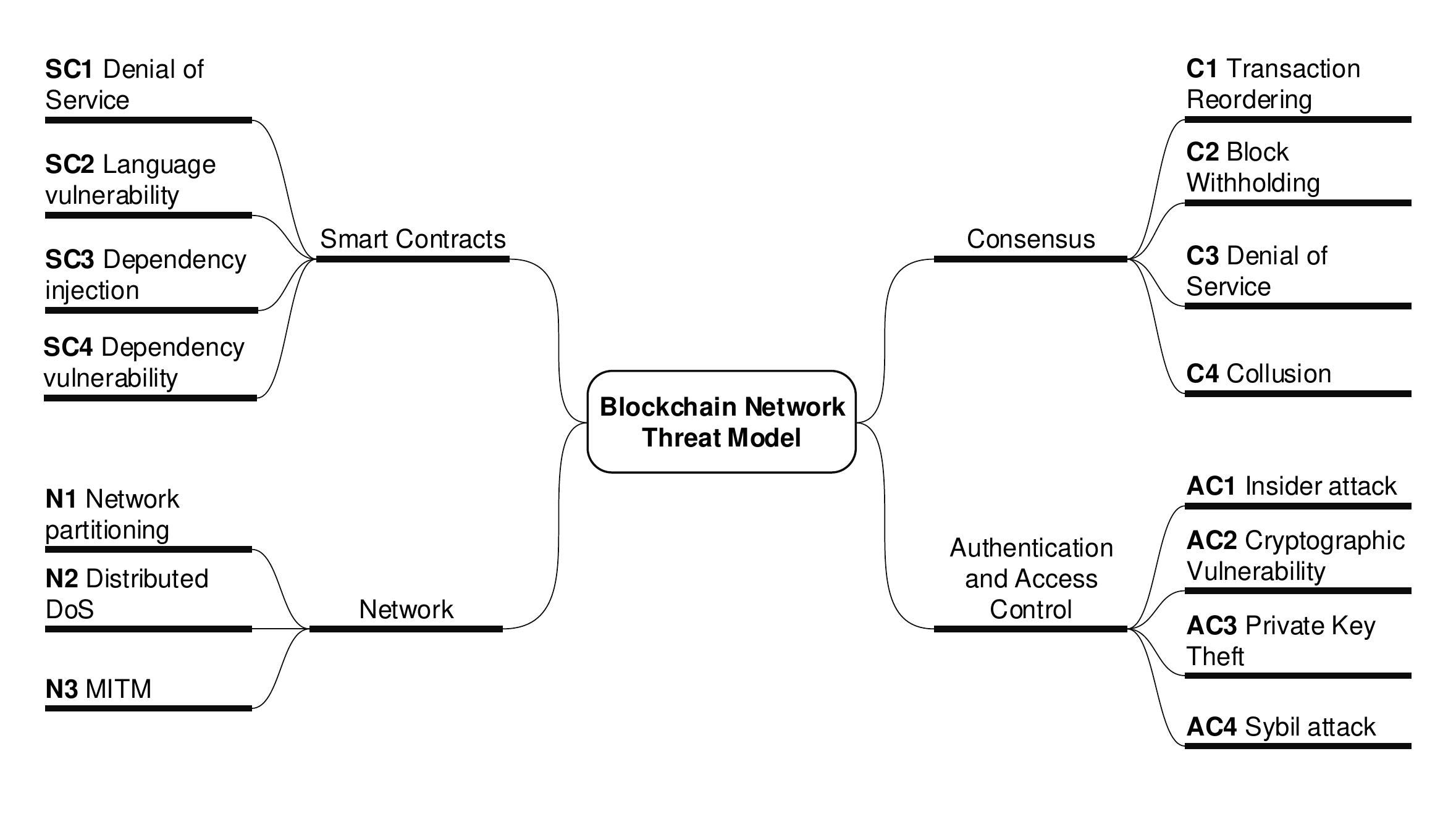}
        \caption{Blockchain Networks Threat Model in attack tree notation.}
        \label{fig:at}
    \end{figure}
    
    To illustrate the monitoring task's complexity, we show an overview of possible attacks in an attack tree notation in Figure \ref{fig:at}. The listed attacks are based on prior work \cite{Putz2020,Putz2019Insider} and related literature surveys \cite{Dabholkar2019,Homoliak2020}. For each leaf on the tree, there are various ways to successfully deploy the attack, which we did not include for conciseness. The attack tree focuses on the blockchain network and nodes. Therefore, it does not include application-specific attacks such as web application vulnerabilities. Each of the shown attacks is indicated by different combinations of threat indicators \cite{Putz2020}. Security experts need to identify threats based on these indicators as part of the \textit{Analysis} process step. Visualizing the indicators provides the necessary overview to identify vulnerable components for in-depth analysis. Therefore, domain experts' overarching task is the \textit{analysis of blockchain data to identify possible threats}, which is to be supported by visualizations. To allow domain experts to execute this work adequately, we have identified more specific tasks based on the attacks and corresponding threat indicators from prior work \cite{Putz2020}. These tasks are shown in Table \ref{tab:tasks}.
    
    \begin{table}[t]
      \centering
      \caption{Security expert tasks and related attacks (cf. Figure \ref{fig:at}).}
        \renewcommand{\arraystretch}{1.15}
        
        \begin{tabular*}{\textwidth}{l @{\extracolsep{\fill}} ll}
        \textbf{Task} & \textbf{Description} & \textbf{Related attacks} \\
        \midrule
        \textit{T1} & Identify vulnerable smart contracts & SC1, SC2, SC3 \\
        \textit{T2} & Identify blockchain framework vulnerabilities & SC4, AC2 \\
        \textit{T3} & Inspect log files of running services on demand & SC4, N1, N3, C3, C4 \\
        \textit{T4} & Review networking activity & N1, N2, N3 \\
        \textit{T5} & Compare transaction metrics over time & N2, C2 \\
        \textit{T6} & Explore block and transaction history & SC1, SC2, C3, AC1 \\
        \textit{T7} & Review configuration changes & C1, AC1 \\
        \textit{T8} & Detect identity abuse & AC1, AC3, AC4 \\
        \end{tabular*}%
        \renewcommand{\arraystretch}{1}
      \label{tab:tasks}%
    \end{table}%
        
    Each task comprises several sub-tasks that help accomplish the main task. To identify vulnerable smart contracts (\textit{T1}), the expert may manually inspect smart contract code or scan smart contracts for vulnerabilities and inspect scan results. Identifying framework vulnerabilities (\textit{T2}) can be accomplished by reading release notes for the framework and its dependencies. Since many anomalies can have multiple causes (i.e., low transaction throughput), log file inspection (\textit{T3}) helps to identify the root cause of anomalies. To review networking activity (\textit{T4}), the main indicators are the count of active connections to other peers, the activity level of those connections, and last seen times of offline peers. Transaction metrics (\textit{T5}) include throughput, latency and unprocessed transactions. Block and transaction history monitoring (\textit{T6}) implies watching the chain of blocks for inconsistencies such as changed blocks or missing transactions. Reviewing configuration changes (\textit{T7}) includes both active and proposed changes to be able to intervene in case of manipulation attempts. Identity abuse (\textit{T8}) is possible during all phases of an identity's lifecycle, so an expert must monitor issuance, usage in transactions, and revocation.
    
    \subsection{Data elements}
    \label{subsec:data}
    Blockchain Frameworks such as Ethereum and Hyperledger Fabric offer a number of data sources for monitoring. The most obvious data sources are blocks and associated transaction data \cite{Tovanich2019}. These can be used to derive active users, smart contracts, and the general level of activity on the network (i.e., transaction throughput). Numerical data on network activity is also provided through metrics, which can be used to raise alerts for anomalous behavior. On a more technical level, each component of the blockchain node also provides log files. These files give detailed information about smart contract execution, consensus protocol violations, and other node internals. They can be helpful to determine the root cause of an anomaly.
    % auf spezifischen Angriff/Gruppe von Angriffen eingrenzen
    
    \subsection{Design Requirements}
    \label{subsec:design_requirements}
    To wrap up this first part of our contribution, we derive the following general requirements for visualizations aiming to support the \textit{Detect} function of the blockchain security monitoring process. The requirements are based on the above user, task, and data characterizations. Although we follow these requirements in the remainder of this work to design our prototype, they can serve as a general collection for respective visualization designs. We summarize the requirements under several main views that a Visual Analytics system supporting the domain experts’ tasks should comprise:
    
    \textbf{R1 - General Security Information:} A view should allow users to overview a series of general, security-relevant information from the monitored blockchain. Attention should be drawn to any changes on the blockchain’s overall configuration (\textit{T7}). Whenever new smart contracts are deployed to the blockchain, they should be checked (automatically or manually) for vulnerabilities. The results of these checks need to be made available for the analysts (\textit{T1}). Additionally, newly discovered vulnerabilities within the applied blockchain framework should be shown to users within this general view (\textit{T2}).
    
    \textbf{R2 - Network View:} Another view should provide access to any data and metrics related to the peers and their network activities. This includes displaying available information about the peers themselves and the respective identities that interact with the blockchain on behalf of the peers (\textit{T8}). This view should also provide visual access to any network-related metrics that assess the overall network’s health (\textit{T4}).
    
    \textbf{R3 - Transaction View:} Domain experts need to access a view displaying information about the blocks and transactions being handled by the blockchain. This includes detailed information on the blocks and transactions themselves (\textit{T6}) as well as a time-based view on transaction-related metrics allowing to identify any changes in typical structure and processing of transactions (\textit{T5}).
    
    \textbf{R4 - Interactivity \& Details:} Any of the previously mentioned views (R1 – R3) needs to be fully interactive to provide the best possible support for domain experts’ tasks and enable exploratory analysis. Whenever suspicious actions or threat indicators are identified, experts also need access to further details and underlying log files (\textit{T3}).

%%%%%%%%%%%%%%%%%%%%%%%%%%%%%%%%%%%%%%% HYPERSEC %%%%%%%%%%%%%%%%%%%%%%%%%%%%%%%%%%%%%%%
\section{HyperSec: Hyperledger Security Monitoring using Visual Analytics}
\label{sec:hypersec}

    We now introduce our prototype \textbf{HyperSec} (\textbf{Hyper}ledger \textbf{Sec}urity Explorer), a modified version of the open-source project Hyperledger Explorer based on the design requirements introduced in Section~\ref{subsec:design_requirements}. The prototype is open-source and available online, along with a demo deployment \footnote{\url{https://github.com/sigma67/hypersec}}. Our modifications address the two remaining layers of the NBGM by designing our solution based on the domain problem and implementating it within a prototype. 

    \subsection{Architecture and Technology}
    \label{subsec:proto_architecture}
\begin{figure}[t]
            \centering
            \includegraphics[trim={0 2cm 0 0},width=\linewidth]{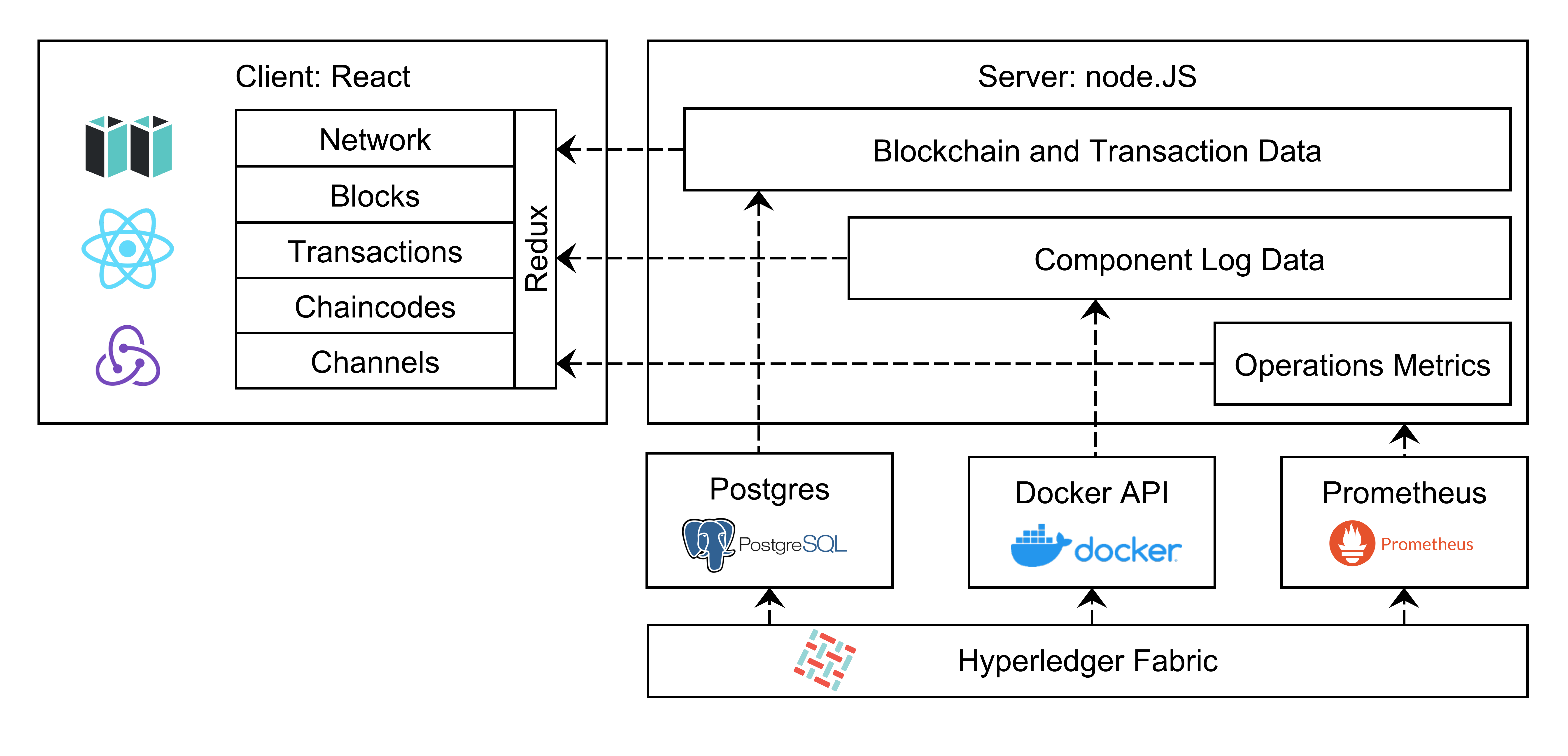}
            \caption{Prototype architecture and data flows.}
            \label{fig:proto_architecture}
        \end{figure}

        We choose Hyperledger Explorer as a starting point since it already provides a working synchronization architecture based on Hyperledger Fabric's block event subscription. We extend the existing architecture to allow for more comprehensive accessibility of relevant data and effective security monitoring. This results in the architecture displayed in Figure \ref{fig:proto_architecture}. We keep the basic structure (data sources, server, and client) of the original architecture for interoperability and transparency reasons. %The necessary changes and extensions were mainly because some data relevant for the monitoring task is not available in the original Hyperledger Explorer architecture. Therefore, 
        However, in our previous study \cite{Putz2020} we found that security-relevant information for Hyperledger Fabric must be retrieved from several data sources: the Hyperledger Fabric SDK, operations metrics, and the application logs available via Docker. Block data is already stored in Hyperledger Explorer's PostgreSQL database. We integrate additional metrics and log sources through server-side proxies to the respective Prometheus and Docker APIs. The React client accesses these through the API exposed by the Hyperledger Explorer server. %Correspondingly, we extend the \textit{Redux} state management layer of the client to be able to stream the new data.
        
        We implement the views defined in Section \ref{subsec:design_requirements} by adapting existing views from the Hyperledger Explorer project. This allows us to retain the frontend structure while introducing new monitoring capabilities. Therefore, domain experts do not need to work with a completely new interface but rather get additional relevant information on the respective views. The updated views host a series of interactive visualizations based on the \textit{visx}\footnote{\url{https://airbnb.io/visx/}} visualization primitives for React. They all follow a similar structure: relevant data is retrieved from the client’s \textit{Redux} state handling, transformed for use in the visual display, mapped into visual primitives, and finally rendered \cite{Chi}.
    
    \subsection{Visual representations and interactions}
    \label{subsec:vis_design}
        We now go into more detail on our HyperSec prototype's visual representations addressing the requirements \textit{R1}-\textit{R3} and their interactivity (\textit{R4}). As mentioned before, we integrate the visualizations into existing Hyperledger Explorer views to retain the familiar structure for domain experts. This Section is structured accordingly to the naming of the original Hyperledger Explorer views.
        
        \begin{figure}[t]
            \centering
            \includegraphics[trim={1cm 1.5cm 1cm 0},width=\linewidth]{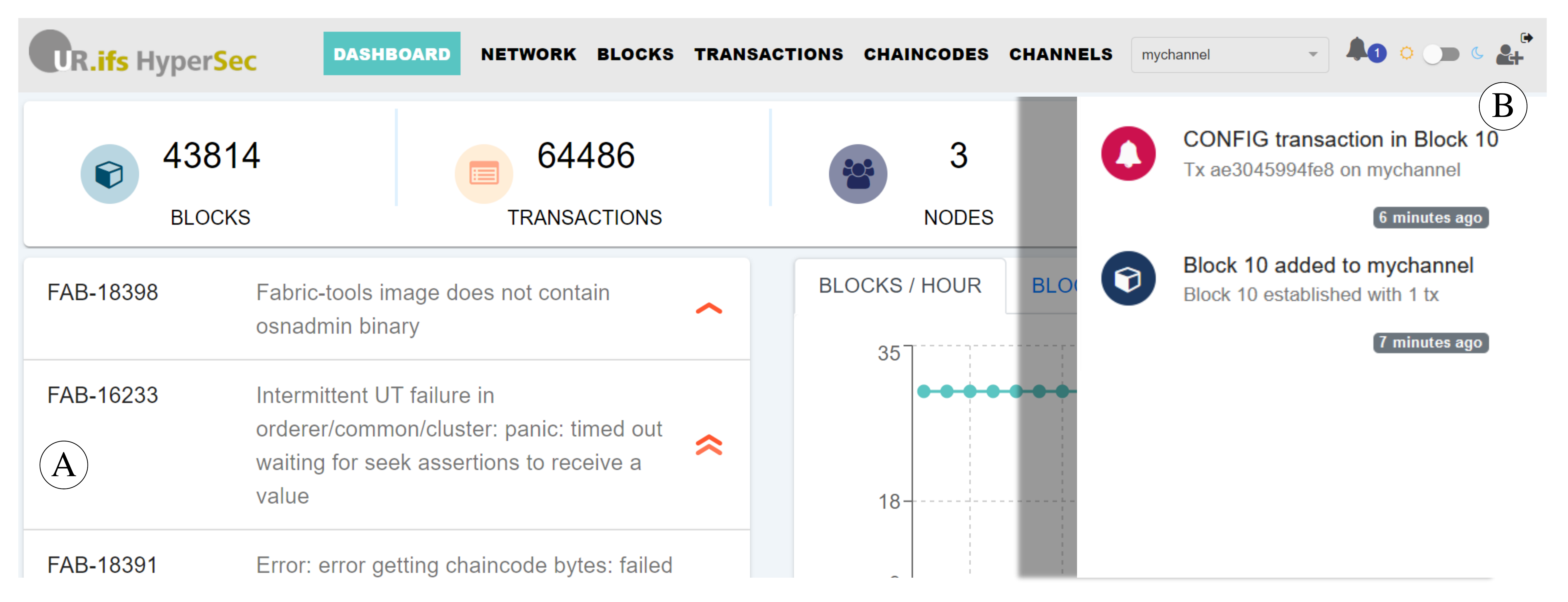}
            \caption{\textit{Dashboard} view: Security issues, alerts and general overview.}
            \label{fig:dashboard}
        \end{figure}
        \textbf{Views \textit{Dashboard} and \textit{Chaincodes}:} To fulfill the Design Requirement \textit{R1}, we adjust two views of the Hyperledger Explorer. First off, directly on the Explorer's landing page, called ``Dashboard'', we show a list of known Hyperledger Fabric issues of High/Highest importance from the Hyperledger JIRA\footnote{\url{https://jira.hyperledger.org}} ordered by last updated (Figure \ref{fig:dashboard}$\mathbf{A}$). Any list item can be expanded to reveal additional information about the issue. Although there is no issue category directly reflecting security issues, this information is highly relevant for \textit{T2 – Identify blockchain framework vulnerabilities}. Additionally, there is no other source for the respective information. In the side menu (\ref{fig:dashboard}$\mathbf{B}$), an alert appears whenever the configuration of the monitored Hyperledger Fabric blockchain is changed (\textit{T7}).
            
        To allow domain experts to detect vulnerable chaincodes, we include available security scans in the respective ``Chaincodes'' view. Whenever a smart contract went through a security scan, analysts can directly check this scan's results in the HyperSec prototype (\textit{T1}). We use the open-source static analysis tool revive-cc\footnote{\url{https://github.com/sivachokkapu/revive-cc}} to detect security vulnerabilities and store the scan result in the Hyperledger Explorer PostgreSQL database. To ensure the scans are up to date, we set up automated jobs to regularly generate security reports of deployed chaincodes.
            
         \begin{figure}[t]
            \centering
            \includegraphics[trim={1cm 1.5cm 1cm 0},width=\linewidth]{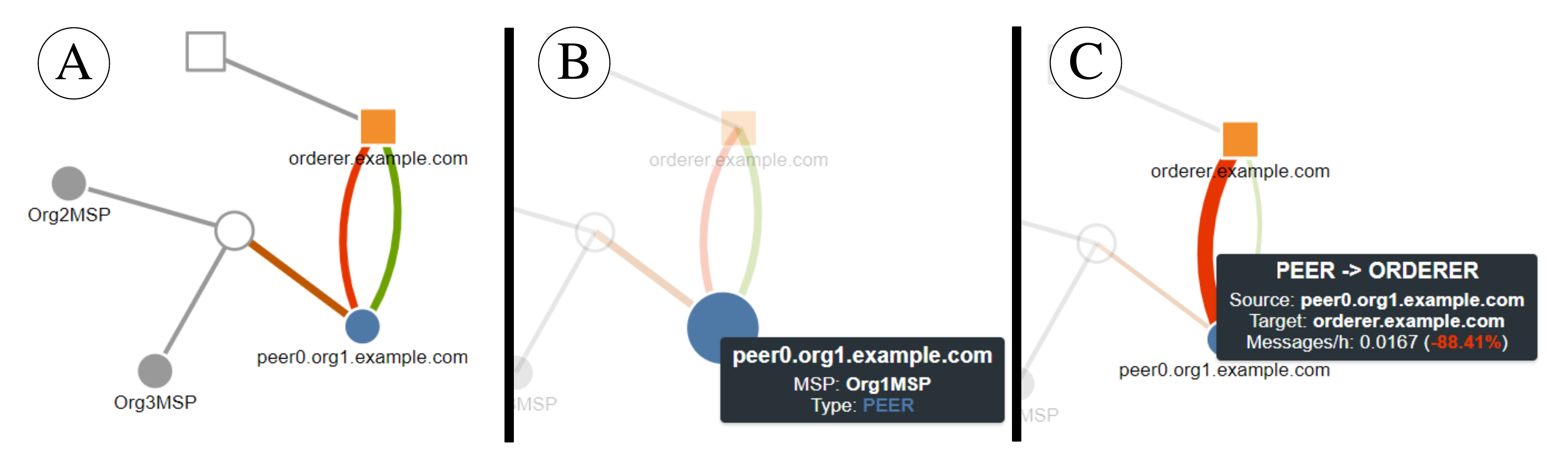}
            \caption{\textit{Network} view: Interactive visualization of network traffic between peers and orderers.}
            \label{fig:nodeLink}
        \end{figure}           
        \textbf{View \textit{Network}:} The \textit{Network} view targets design requirement \textit{R2} intended for tasks \textit{T4} and \textit{T8}. The original Hyperledger Explorer shows a tabular list with basic information about the peers connected to the monitored Hyperledger Fabric network. In our HyperSec prototype, we extend this table with a force-directed node-link diagram to effectively visualize networking activity (Figure \ref{fig:nodeLink}$\mathbf{A}$). The nodes' different shapes indicate different peer types within the network: Circles are used to display peers while rectangles represent orderers. Links between the glyphs are used to display known networking activities. 
        
        However, the unavailability of core information restricts this view's expressivity. While rich information about the peers can be easily retrieved from the Hyperledger Explorer, no data about the peers' network connections is provided. Therefore, HyperSec retrieves networking information directly from Hyperledger Fabric through the Prometheus API. By doing so, experts get at least some information about the peers' networking activity within the own Membership Service Provider (MSP). However, as the Hyperledger Fabric network is decentralized, it is not possible to get any information about other MSPs' networking activities. Because of this restriction, we introduce two empty nodes in the node-link diagram (uncolored nodes in Figure \ref{fig:nodeLink}$\mathbf{A}$), which mark the border of the monitoring visibility regarding networking activities. Nodes within the owned MSP are colored; those within other MSPs are greyed out.
        
        Links connecting the nodes in the graph represent known network connections. Again, outside the own MSP's borders, expoerts do not get much information. Therefore, we connect any foreign peer and orderer to the respective artificial node. The coloring of the links follows a continuous scale from -1 to 1. This scale measures the current deviation of the link's message traffic from the average, by comparing traffic in the last hour with traffic in the previous seven days. If this deviation is low, the link is colored in a green tone. A red link, in contrast, marks a high variation of message numbers.
        
        The node-link diagram is fully interactive. Nodes are draggable to ensure that security analysts can adjust the layout to their own needs if necessary. Hovering over nodes (Figure \ref{fig:nodeLink}$\mathbf{B}$) or links (Figure \ref{fig:nodeLink}$\mathbf{C}$) highlights the hovered object and shows additional status information about it.
        
        \begin{figure}[t]
            \centering
            \includegraphics[trim={1cm 1.5cm 1cm 0},width=\linewidth]{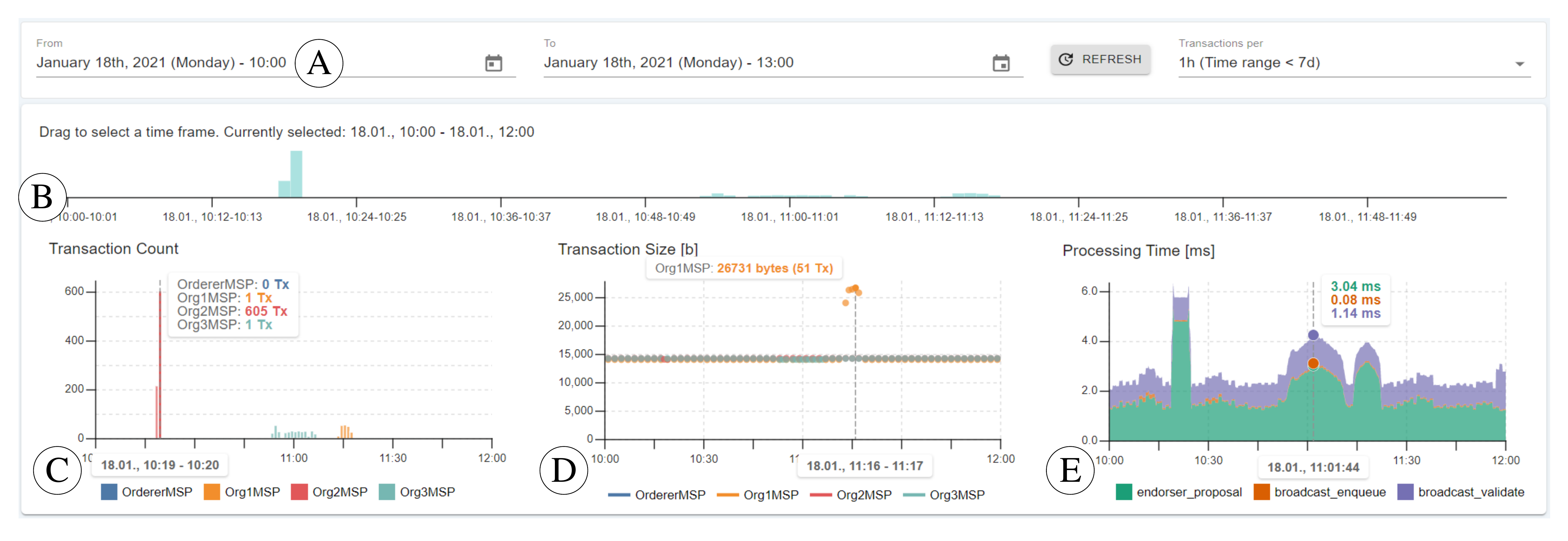}
            \caption{\textit{Transactions} view: Interactive visualizations for transaction count, size and processing time.}
            \label{fig:transactionView}
        \end{figure}
        \textbf{View \textit{Transactions}:} This view (\textit{R3}) satisfies tasks \textit{T5} and \textit{T6}. Some modifications to the original simple table view ensure that the transactor identity and transaction size are visible. The primary adjustment we made to this view is introducing four visualizations (Figure \ref{fig:transactionView}). 

        We make small adjustments to the original timeframe selection (Figure \ref{fig:transactionView}$\mathbf{A}$). The selection defines the time range for which information about transactions should be displayed. On the right side of the timeframe selection, we added a dropdown menu to select the aggregation granularity (1 minute, 1 hour, 12 hours, 24 hours) for the visualizations. This helps security experts if they need to compare and contextualize available information.

        The wide bar chart (Figure \ref{fig:transactionView}$\mathbf{B}$) always displays the entire selected date range. Each bar represents the number of transactions within a specific range of time specified through the aggregation granularity. This bar chart supports analysts in navigating the selected time range. A brushing interaction (horizontal dragging on the chart) selects an even smaller time range for detailed analysis. On interaction, the other visualizations (Figure \ref{fig:transactionView}$\mathbf{C}$, $\mathbf{D}$, and $\mathbf{E}$) and the transactions table are dynamically updated with data from this narrowed time range.

         A stacked bar chart (Figure \ref{fig:transactionView}$\mathbf{C}$) visualizes the number of transactions per aggregation window. However, it does this only for the transactions selected through the brushing interaction on the visualization \ref{fig:transactionView}$\mathbf{B}$. It shows the transaction count's composition based on which MSP contributed how many transactions. The scatterplot \ref{fig:transactionView}$\mathbf{D}$ shows the transaction size in bytes throughout the time range for each MSP. Each circle on the scatterplot represents the average size of transactions submitted by a specific MSP. This aggregation is performed to scale the chart for large numbers of transactions. During attacks, thousands of transactions can be submitted within just minutes, thus freezing the chart if each transaction were drawn individually. The stacked area chart \ref{fig:transactionView}$\mathbf{E}$ finally shows the development of three different metrics, which we identify as helpful to get an idea for the processing time in seconds that a transaction needs from proposal to validation. As information for processing times are not available distinctively per transaction but continuously per time unit, we choose to display this metrics with a continuous visualization technique.

        The visualizations \ref{fig:transactionView}$\mathbf{C}$, $\mathbf{D}$, and $\mathbf{E}$ are again fully interactive. Hovering individual bars or hovering along the continuous sizes and times displays additional information as tooltips. Different metrics can also be toggled using the legend icons below the visual representations.
%notes: 
%- block has no timestamp in fabric (open issue in JIRA)
%- 

%%%%%%%%%%%%%%%%%%%%%%%%%%%%%%%%%%%%%%% EVALUATION %%%%%%%%%%%%%%%%%%%%%%%%%%%%%%%%%%%%%%%
\section{Evaluation}
\label{sec:use_case}

For our evaluation, we focus on three common attacks that cover the majority of the tasks outlined in Table \ref{tab:tasks}: \textbf{SC2}, \textbf{N2}, and \textbf{AC1}. We simulate these attacks a Hyperledger Fabric test network, which the HyperSec prototype is connected to.

\textbf{SC2} refers to a language vulnerability, i.e. a software bug that exposes chaincode to malicious exploits. A security expert may become aware of such an exploit by identifying vulnerable smart contracts (\textit{T1}) and by inspecting transaction history (\textit{T6}). For example, consider a read-after-write vulnerability detected by the chaincode scanner \textit{revive-cc}. The security expert can inspect an automatically generated chaincode scan in the \textit{Chaincodes} view. Intuitively, the experts check for past exploitations using the \textit{Transactions} view. Thereto, the transactions table can be filtered using the chaincode name and applicable time frame. The filtered transactions can be inspected individually to find unusual read/write sets. 

\textbf{N2} refers to a distributed denial of service attack. If a peer or orderer is targeted by a traffic-based denial of service attack, its connection to other peers will be impaired as well. The \textit{Network} view (Figure \ref{fig:nodeLink}$\mathbf{C}$) shows high deviation in gossip communication traffic to the targeted peer during such an attack (\textit{T4}). If the local peer is targeted, the metrics in the \textit{Transaction} view (Figure \ref{fig:transactionView}$\mathbf{E}$) show increased transaction processing latency due to high peer load (\textit{T5}). For attackers that can send transaction to the network, transaction-based DoS is more effective. Figure \ref{fig:transactionView}$\mathbf{C}$ and \ref{fig:transactionView}$\mathbf{D}$ show two such attempts using high transaction volume ($\mathbf{C}$) and large transaction size ($\mathbf{D})$. \ref{fig:transactionView}$\mathbf{E}$ also shows spikes in processing latency during the time of attack (spikes 1 and 3 in that chart).

To investigate the source of the anomaly, experts can check the peer logs, which are available in the Network view (\textit{T3}). They cross-reference any error messages with open issues in the Hyperledger JIRA, which are available in the Dashboard view (\textit{T2}).

\textbf{AC1} refers to an attack where an insider abuses valid credentials for malicious purposes. Consider an insider attempting to corrupt the blockchain network's configuration using a configuration transaction. Security experts are immediately notified about the configuration change in the notification sidebar (\textit{T7}, see Figure \ref{fig:dashboard}). Details of the attempted configuration change are available in the transaction history table (\textit{T6}), where the full read-write set of the transaction is available by selecting the respective transaction.

%%%%%%%%%%%%%%%%%%%%%%%%%%%%%%%%%%%%%%% CONCLUSION %%%%%%%%%%%%%%%%%%%%%%%%%%%%%%%%%%%%%%%
\section{Conclusion}
\label{sec:conclusion}
This work introduced the task-oriented design and prototypical implementation of HyperSec, a visual analytics security monitoring tool tailored for Hyperledger Fabric. Throughout the design of HyperSec, we followed the NBGM design methodology. The domain problem describes the activities of the blockchain security monitoring process to be supported by visualizations. Subsequently, we identified the involved users, their specific tasks, and the available data elements. These considerations culminated in design requirements that apply to any visualization system aiming to support blockchain security analysts. Our prototype HyperSec picks up on these design requirements. It extends the open-source architecture of Hyperledger Explorer with additional security-relevant data sources. The data is aggregated, processed and displayed in appropriate visualizations supporting blockchain security analysts to detect potential attacks.

Our prototype might not cover every possible subtask of the defined tasks of blockchain security analysts. This is in part due to limited availability of data provided by Hyperledger Fabric itself. We plan to update our prototype as additional data sources become available in the future, and are open to contributions from the community. 

The security of the monitoring tool itself is also important, as it should not contribute additional attack vectors by leaking blockchain data. During our implementation we found some bugs and vulnerabilities within Hyperledger Explorer, which we subsequently fixed and contributed to the upstream project.

%Future Work
%manual or automated transaction anomaly detection based on data stored in transaction database (anomaly detection requires data preprocessing, and HL Explorer already provides a suitable database)

%integration of blockchain-specific threat intelligence data (z.B. CVEs o.ä. mit Bezug auf DLT?)

\bibliographystyle{splncs04}
\bibliography{references}

\end{document}